\documentclass[12pt]{article}

\usepackage{epsf}

\newcommand{\SDVA}{\langle S_2 \rangle}

\begin{document}

\begin{center}
{\bfseries AZIMUTHAL STRUCTURES OF PRODUCED PARTICLES\\
 IN HEAVY ION INTERACTIONS}

\vskip 5mm

S.~Vok\'al$^{1 \dag}$, A.~Krav\v c\'akov\'a$^{2}$, S.~Lehock\'a$^{2}$ and G.~I.~Orlova$^{3}$

\vskip 5mm

{\small
(1) {\it
VBLHE, JINR, Dubna, Russia
}
\\
(2) {\it
University of  P.~J.~\v{S}af\'arik, Ko\v{s}ice, Slovakia
}
\\
(3) {\it
Lebedev Physical Institute, Moscow, Russia
}
\\
$\dag$ {\it
E-mail: vokal@sunhe.jinr.ru
}}
\end{center}

\vskip 5mm

\begin{center}
\begin{minipage}{150mm}
\centerline{\bf Abstract}
The angular substructures of particles produced in ${}^{208}$Pb at 158~A~GeV/c and ${}^{197}$Au at 11.6~A~GeV/c induced interactions with Ag(Br) nuclei in emulsion detector have been investigated. Nonstatistical ring-like substructures of produced particles in azimuthal plane of a collision have been found and their parameters have been determined. 
\end{minipage}
\end{center}

\vskip 10mm


\section{Introduction}

An important aim of nucleus collision investigation at high energies is to search for a phenomena connecting with large densities obtained in such collisions. As an example, the transition from the QGP (quark-gluon plasma) back to the normal hadronic phase is predicted to give large fluctuations in the number of produced particles in local regions of phase space \cite{bib01,bib02}. The observed effects of such type are dominated by statistical fluctuations. Significant deviations from them are only observed after removing the statistical part of the fluctuations \cite{bib03}.

In case of angular structures of produced particles investigation two different classes were revealed, which could be referred to as jet-like and ring-like substructures.

The goal of our work was to study the ring-like substructures of produced particles in azimuthal plane. They are occurrences if many particles are produced in a narrow region along the rapidity axis, which at the same time are diluted over the whole azimuth. The jet-like substructures consist of cases where particles are focused in both dimensions \cite{bib04}. \\

For the first time the individual nucleus-nucleus collisions with a ring-like substructure of produced particles in the azimuthal plane have been observed more then 20 years ago in cosmic ray experiments \cite{bib05}. Later a lot of the nucleus-nucleus collisions with the ring-like substructure were observed in the accelerator experiments at high energy \cite{bib03,bib06,bib07,bib08,bib19,bib20}.

A new mechanism of multiparticle production at high energies was proposed in \cite{bib09,bib10,bib11}. This mechanism is similar to that of Cherenkov electromagnetic radiation. As a hadronic analogue, one may treat an impinging nucleus as a bunch of confined quarks each of which can emit gluons when traversing a target nucleus \cite{bib12,bib13}. The idea about possible Cherenkov gluons is relying \cite{bib09} on experimental observation of the positive real part of the elastic forward scattering amplitude of all hadronic processes at high energies. This is a necessary condition for such process because in the commonly used formula for the refractivity index its excess over 1 is proportional to this real part. Later I.~M.~Dremin \cite{bib10} noticed that for such thin targets as nuclei the similar effect can appear due to small confinement length thus giving us a new tool for its estimate. If the number of emitted gluons is large enough and each of them generates a mini-jet, the ring-like substructure will be observed in the target (azimuthal) diagram. If the number of emitted gluons is not large, we will see several jets correlated in their polar, but not in the azimuthal angles. Central collisions of nuclei are preferred for observation of such effects because of a large number of participating partons.

In the present study the ring-like substructures of charged produced particles from ${}^{208}$Pb and ${}^{197}$Au induced interactions with Ag(Br) target nuclei in emulsion detector at 158~A~GeV/c and 11.6~A~GeV/c, correspondently, have been analyzed. The comparison with the FRITIOF calculations \cite{bib14} has been made. All used data are obtained in the frames of EMU01 Collaboration.


\section{Experiment}

The stacks of NIKFI~BR-2 nuclear photoemulsions have been irradiated horizontally by ${}^{208}$Pb beam at 158~A~GeV/c (the CERN SPS accelerator -- experiment EMU12) and by ${}^{197}$Au beam at 11.6~A~GeV/c (the BNL AGS accelerator -- experiment E863). 

The photoemulsion method allows to measure:
      \textit{multiplicities of any charged particles}: 
      produced particles ($N_s$) with $\beta > 0.7$, 
      projectile fragments ($N_F$) with $\beta \approx 0.99$ 
      and target fragments ($N_h$) with $\beta < 0.7$,
      \textit{angles of particles} with the resolution of $\Delta\eta = 0.010-0.015$ 
      rapidity units in the central region, pseudo-rapidity 
      is given by $\eta = -\ln(\tan(\theta/2))$, and $\theta$ is the emission angle 
      with respect to the beam direction,
      \textit{charges of projectile fragments} $Z_F$.

Further details on both experiments, measurements and experimental criteria can be found in \cite{bib15,bib16}.

In this work we have analyzed:
\begin{itemize}
\item 628 Pb+Ag(Br) collisions found by the along-the track scanning. From the collisions we have 
      selected three centrality groups determined by the multiplicity of the produced particles:
      $350 \leq N_s < 700, 700 \leq N_s < 1000$ and $N_s \geq 1000$. As it was shown in our
      previous paper \cite{bib18} the criterion $N_s \geq 350$ selects the interactions 
      of lead nuclei at 158~A~GeV/c with the heavy emulsion targets Ag and Br with $b_{imp} < 8
      $~fm only. Moreover the group with $N_s \geq 1000$ comprises the central Pb+Ag(Br)
      interactions with impact parameter $b_{imp} \approx (0 - 2)$~fm.

\item 1128 Au+Ag(Br) collisions found by the along-the track scanning. From the collisions we have
      selected analogous three centrality groups determined by the multiplicity of the produced
      particles: $100 \leq N_s < 200, 200 \leq N_s < 300$ and $N_s \geq 300$.
\end{itemize}


\section{Method}

A method we use to search for a ring-like substructure and to determine parameters they have been devised in paper \cite{bib03}. The produced particle multiplicity $N_d$ of analyzed subgroup from an individual event is kept a fixed. Each $N_d$-tuple of consecutive particles along the $\eta$-axis of individual event can then be considered as a subgroup characterized by 
\textit{a size}: $\Delta\eta = \eta_{max} - \eta_{min}$, where $\eta_{min}$ and $\eta_{max}$ are the pseudo-rapidity values of the first and last particles in the subgroup, 
by \textit{a density}: $\rho_d = \frac{N_d}{\Delta\eta}$ and 
by \textit{a average pseudo-rapidity} (or a subgroup position): $\eta_m = \frac{\sum\eta_i}{N_d}$.
Another way is to kept a fixed the $\Delta\eta$ interval. This method has been used by G.~L.~Gogiberidze et al. in papers \cite{bib19,bib20}.

To parameterize the azimuthal structure of the subgroup in a suitable way a parameter of the
azimuthal structure $S_2 = \sum(\Delta\Phi_i)^2$ 
have been suggested, where $\Delta\Phi$ is the difference between azimuthal angels of two neighboring particles in the investigated subgroup (starting from the first and second and ending from the last and first). For the sake of simplicity it was counted $\Delta\Phi$ in units of full revolutions $\sum(\Delta\Phi_i) = 1$. 

The parameter $S_2$ is large $(S_2 \rightarrow 1)$ for the particle groups with the jet-like structure and is small $(S_2 \rightarrow 1/N_d)$ for the particle groups with the ring-like structure. The expectation value for the parameter $S_2$, in the case of stochastic scenario with independent particles in the investigated group, can be analytically expressed as $\SDVA = \frac{2}{N_d + 1}$ 
This expectation value can be derived from the distribution of gaps between neighbors.

What can one wait to see in the experimental $S_2$ -- distributions in different scenarios? As it was shown in \cite{bib18} in case of a pure stochastic scenario the normalized $S_2/\SDVA$ -- distribution would have peak position at $S_2/\SDVA = 1$. The existence of the jet-like substructures in collisions results to appearance of additional  $S_2/\SDVA$ -- distribution from this effect but shifted to the right side in comparison with stochastic distribution. Analogously, the existence of the ring-like substructures results to appearance of additional $S_2/\SDVA$ -- distribution from this effect but shifted to the left side. As result, the summary $S_2/\SDVA$ -- distribution from this three effects may have different form depends of mutual order and sizes \cite{bib18}.


\section{Results}

The first detailed study of the average values of the parameter $S_2$ was performed in \cite{bib03}. The azimuthal substructures of particles produced within dense and dilute groups along the rapidity axis in the central ${}^{16}$O and ${}^{32}$S induced collisions with Ag(Br) and Au targets at 200~A~GeV/c (EMU01 data sets) were analyzed. The results were compared with different FRITIOF calculations including $\gamma$-conversion and the HBT effects. It was conclude that jet-like and ring-like events do not exhibit significant deviations from what can be expected from stochastic emission.

The study of the $S_2$-parameter distributions for subgroups of the particles produced in ${}^{197}$Au interactions at 11.6~A~GeV/c with Ag(Br) targets in emulsion detector has been done in \cite{bib17}. Nonstatistical ring-like substructures have been found and cone emission angles as well as other parameters they have determined. 

\begin{figure}[!h]
\begin{center}
\epsffile{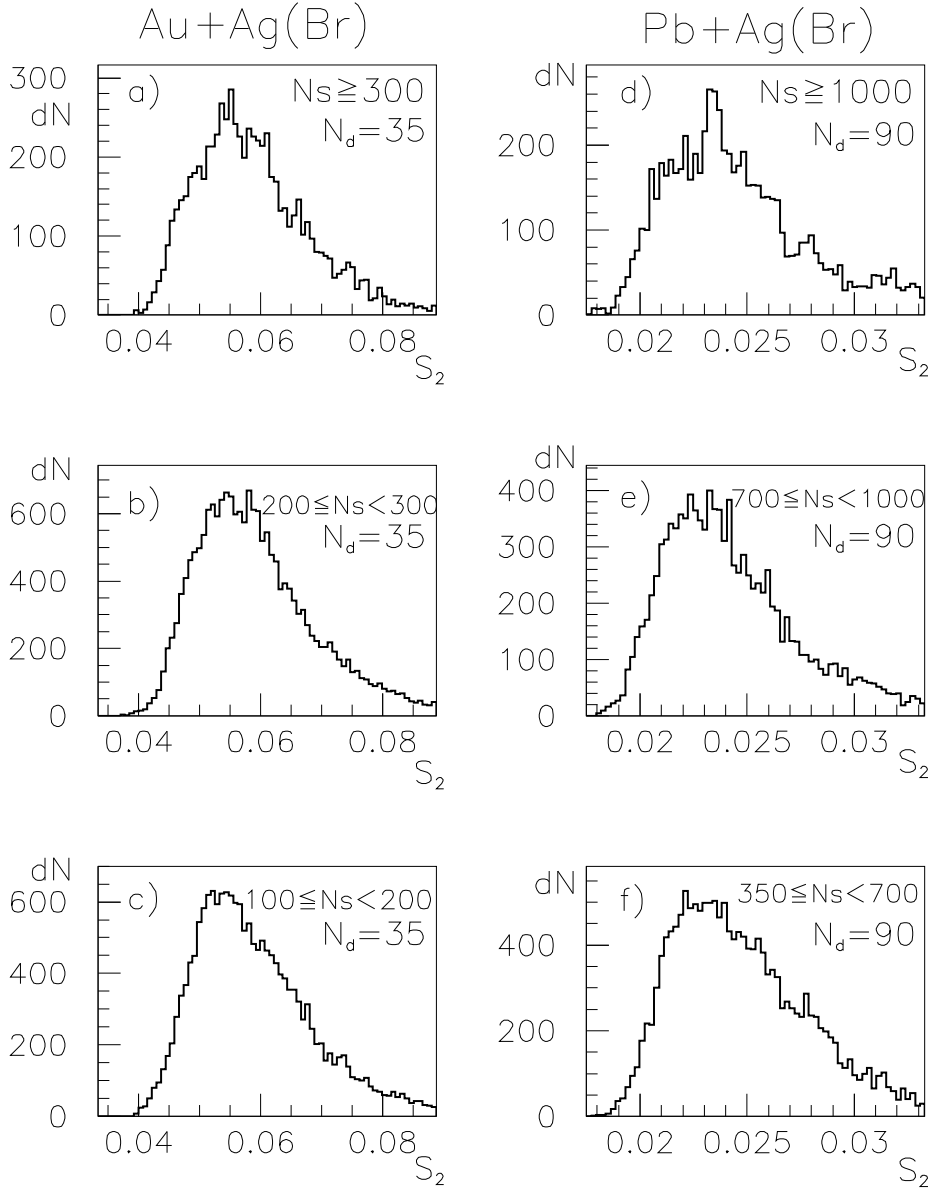}
\end{center}
\caption[*]{ (a, b, c) The $S_2$ -- distributions for subgroups with $N_d = 35$ and different groups of shower particles multiplicity $N_s$ in Au+Ag(Br) collisions at 11.6~A~GeV/c;\\
(d, e, f)  The $S_2$ -- distributions for subgroups with $N_d = 90$ and different groups of shower particles multiplicity $N_s$ in Pb+Ag(Br) collisions at 158~A~GeV/c. }
\label{fig04}
\end{figure}

In Fig.~\ref{fig04}(a -- c) the $S_2$ -- distributions for groups with $N_d = 35$ are shown for Au+Ag(Br) collisions with multiplicities of the produced particles $N_s > 300\; (a),\; 200 \leq N_s < 300\; (b)$ and $100 \leq N_s < 200\; (c)$. The analogical $S_2$ spectra for subgroups with $N_d = 90$ obtained in Pb+Ag(Br) collisions are shown in Fig.~\ref{fig04}(d -- f) for different centrality groups with $N_s > 1000\; (d),\; 700 \leq N_s < 1000\; (e)$ and $350 \leq N_s < 700\; (f)$. As one can see at all three cases of different centralities the $S_2$ -- distributions have the peak position around the value corresponding to the stochastic scenario $(S_2/\SDVA = 1)$ and tails at the right side. In order to study the ring-like substructures the only left part of the $S_2$ -- distribution, where a signal of ring-like substructure may be expected, is essential. As one can see the only central collisions $(N_s > 300$ in Au- and $N_s > 1000$ in Pb-induced collisions with Ag(Br) targets) have a proved additional peak on the left part. This indicates that a certain ring-like substructures are present at these two experiments.\\

\begin{figure}[!t]
\epsfysize=7cm
\begin{center}
\epsffile{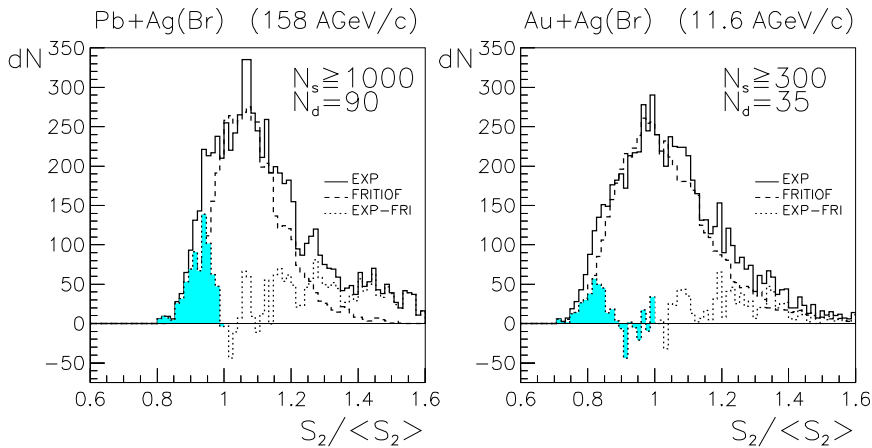}
\end{center}
\caption[*]{ The experimental and FRITIOF model normalized $S_2/\SDVA$ -- distributions for central ${}^{208}$Pb and ${}^{197}$Au induced collisions with Ag(Br) targets in the emulsion detector. Here, $N_s$ is the number of produced particles and $N_d$ is the number of particles in the analysed subgroup. }
\label{fig06}
\end{figure}

The experimental normalized $S_2/\SDVA$ -- distributions compared with the calculated ones by the FRITIOF model for the most central groups of events measured in ${}^{197}$Au and ${}^{208}$Pb induced collisions with Ag(Br) nuclei at 11.6 and 158~A~GeV/c are presented on the Fig.~\ref{fig06}. The model distributions were aligned according to the position of the peak with the expe\-ri\-mental one. The FRITIOF model includes neither the ring-like nor the jet-like effects, so the model distributions are used like the statistical background.\\

One can see that both experimental distributions are shifted to the right, have a tail in the right part and are broader than the spectra calculated by the FRITIOF. The left parts of both experimental distributions are not as smooth as in the model and there are some shoulders that refer to the surplus of the events in this region. 

The results obtained from the experimental data after the subtraction of the statistical background are also shown on this figure. The resultant distributions have two very good distinguishable hills, the first in the region $S_2/\SDVA < 1$, where the ring-like effects are expected and the second in the jet-like region --  $S_2/\SDVA > 1$. The probability of the formation of the nonstatistical ring-like substructures can be estimated as a rate of the surface of the ring-like part to the full surface of the experimental distribution.\\

Our preliminary results for ${}^{208}$Pb+Ag(Br) collisions at 158~A~GeV/c are shown that the estimated contribution of the events with  nonstatistical ring-like substructures in the emission of produced particles is about $10 - 12 \%$ in the most central group of collisions with $N_s \geq 1000$. This value slowly decreases in two other groups of less central events with $350 \leq N_s < 700$ and $700 \leq N_s < 1000$.\\

\begin{figure}[!t]
\epsfysize=7cm
\begin{center}
\epsffile{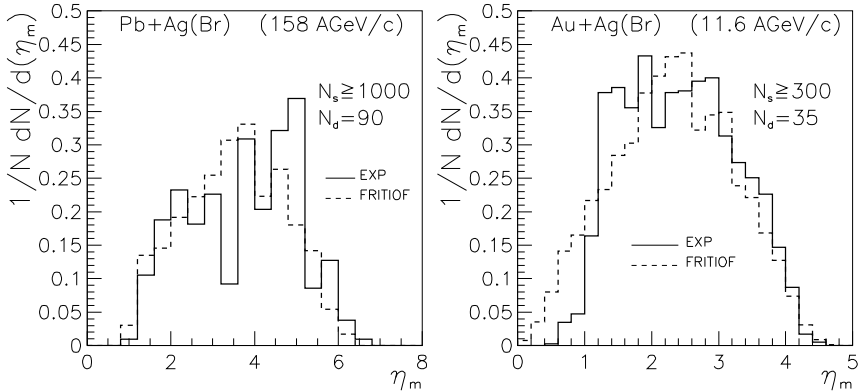}
\end{center}
\caption[*]{ The ring-like subgroup position $(\eta_m)$ distributions for central experimental (solid histogram) and FRITIOF model (dashed histogram) for ${}^{208}$Pb+Ag(Br) and ${}^{197}$Au+Ag(Br) collisions. }
\label{fig07}
\end{figure}

To analyze the ring-like subgroup position on the pseudorapidity axis the $\eta_m$ -- distributions for subgroups with $S_2/\SDVA < 1$ from central collisions are presented for experimental data and FRITIOF model in Fig.~\ref{fig07}. 
One can see that the experimental distributions have a downfall in the central region, where the produced particle and FRITIOF distributions have maximum and two symmetrical hills from both sides of the center. For central ${}^{208}$Pb+Ag(Br) collisions $\eta_m$ -- distribution has two hills  -- one at $\eta = 1.6 - 3.2$ and other at $\eta = 3.6 - 5.2$, the center of the distribution for produced particle is at $\eta \approx 3.5$. For central ${}^{197}$Au+Ag(Br) collisions $\eta_m$ -- distribution has two hills -- one at $\eta = 1.2 - 2.0$ and other at $\eta = 2.2 - 3.0$, the center of the distribution is at $\eta \approx 2.2$. The downfall of the $\eta_m$ -- distributions is more visible for ${}^{208}$Pb+Ag(Br) interactions that are probably connected with larger cross section of the effect for the collisions with bigger multiplicity that realized at higher beam energy and for more central collisions.\\

\begin{figure}[!h]
\epsfysize=7cm
\begin{center}
\epsffile{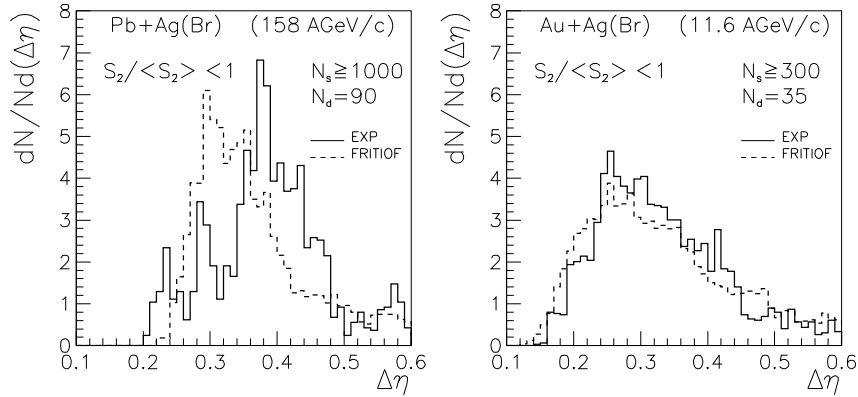}
\end{center}
\caption[*]{ Comparison of the experimental and the FRITIOF model $\Delta\eta$ -- distributions for central interactions of ${}^{208}$Pb and ${}^{197}$Au nuclei with Ag(Br) targets for ring-like region ($S_2/\SDVA < 1$). }
\label{fig09}
\end{figure}

To investigate the ring-like subgroups size $\Delta\eta$ in Fig.~\ref{fig09} the $\Delta\eta$ -- distributions are shown in the region of the ring-like effects $(S_2/\SDVA < 1)$ for the most central ${}^{208}$Pb $(N_s \geq 1000, N_d = 90)$ and ${}^{197}$Au $(N_s \geq 300, N_d = 35)$ induced collisions with Ag(Br) targets compared with FRITIOF model. One can see that there are some distinctions in the shapes of the experimental and model distributions for Pb induced collisions. There appeared 3 or 4 peaks in the experimental $\Delta\eta$ -- distributions in the ring-like effect region that we don't see in other cases. The difference for ${}^{197}$Au data is not so obvious. Moreover, in our previous paper \cite{bib18} it was shown that from one side there are some distinctions in the shapes of the experimental distributions for the regions $S_2/\SDVA < 1$ and $S_2/\SDVA > 1$ but from the other side there are no differences in the $\Delta\eta$ -- distributions calculated by the model for both classes of events ($S_2/\SDVA < 1$ and $S_2/\SDVA > 1$).\\

If the ring-like substructures have been appeared due to an effect analogous to Cherenkov light there may be in a collision two such substructures forming two produced particle cones -- one in the forward and another in the backward direction in center-of-mass system. In such case the cones must have the equal emission angles, because as well known the Cherenkov emission angel depends on the refractive index of matter only. In our case, in case of nuclear matter, it is a way to measure the refractive index of nuclear matter. It is interesting to note that the refractive index of nuclear matter has to be changed in the case of the changes the nuclear matter properties, for example, in the case of phase transition from a normal hadronic matter to quark-qluon plasma.


\section{Conclusion}

The azimuthal ring-like substructures of produced particles from collisions induced by the 11.6~A~GeV/c ${}^{197}$Au and 158~A~GeV/c ${}^{208}$Pb beams with Ag(Br) targets in the emulsion detector have been investigated.
\begin{itemize}
\item The additional subgroups of produced particles in the region of the ring-like substructures  
      ($S_2/\SDVA < 1$) in comparison to the FRITIOF model calculations have been observed.
\item The difference with the FRITIOF model calculations in the $\eta_m$ -- distributions in ring-
      like region $S_2/\SDVA < 1$ indicates to existence of two symmetrical $\eta_m$ -- regions of 
      preferred emission of ring-like substructures -- one in the forward and second in the 
      backward direction in center-of-mass system.
\item The $\Delta\eta$ -- distribution, which gives the information about a ring-like substructure 
      size in pseudorapidity scale, for the experimental data in ring-like region ($S_2/\SDVA < 1
      $) differs from the FRITIOF model calculations.
\item The nonstatistical ring-like substructures formation is more visible for central collisions 
      and for bigger energies. 
\item The results are in good agreement with an idea that the ring-like substructures have been 
      appeared due to an effect analogous to Cherenkov light.
\end{itemize}


\section*{Acknowledgements}

This work was supported by Science and Technology Assistance Agency (Slovak Republic) under the contract No. APVT-51-010002 and by RFBR grants No. 03-02-17079 and No. 04-02-16593.

\end{document}